\documentclass{aastex631}
\usepackage{microtype}
\usepackage{multirow}
\usepackage{gensymb}
\usepackage{xcolor}
\usepackage{amsmath}
\usepackage[bottom]{footmisc}
\usepackage{comment}
\usepackage{float}
\usepackage{xcolor}
\usepackage{placeins}
\usepackage{etoolbox}
\patchcmd{\endthebibliography}{\newpage}{}{}{}

\usepackage{graphicx}
\usepackage{amsmath}
\usepackage{amssymb}
\usepackage{booktabs}

\newcommand{\rhostar}{\rho^{*}}

\newcommand{\dd}{\mathrm{d}}

\begin{document}

\title{An improved self-gravity treatment in general relativistic smoothed particle hydrodynamics}

\author{Pritam Banerjee}
\affiliation{Indian Institute of Astrophysics, II Block, Koramangala, Bengaluru 560034, India}
\email{pritam.banerjee@iiap.res.in}

\author{Kowsona Chakraborty}
\affiliation{Department of Physics, Indian Institute of Technology Kanpur, Kanpur 208016, India}
\email{kowsonac24@iitk.ac.in}

\author{Niles Mondal}
\affiliation{Department of Physics, Indian Institute of Technology Kanpur, Kanpur 208016, India}
\email{nilesm24@iitk.ac.in}

\author{Tapobrata Sarkar}
\affiliation{Department of Physics, Indian Institute of Technology Kanpur, Kanpur 208016, India}
\email{tapo@iitk.ac.in}

\begin{abstract}
General relativistic smoothed particle hydrodynamics (GRSPH) on a fixed black hole (BH) spacetime is suitable for fluid bodies that are much less massive than the BH, but the self-gravity of the fluid is added as an extra term in the equation of motion. Standard GRSPH implementations add this as a Newtonian source term to the momentum equation, which is no longer valid when the fluid moves at a sizeable fraction of the speed of light, close to the BH. We develop an improved prescription by deriving the weak gravitational field of each fluid element in its local rest frame and transforming it to the global frame where the momentum equation is solved. 
The resulting force uses rest-frame distances and softening lengths, accounts for the time dilation of the receiving particle, and its motion relative to the source particle. 
The derivation applies to any weakly self-gravitating flow on a fixed BH background, provided the body is small compared with its distance from the BH. Implemented in a tree-based GRSPH code, the prescription reproduces the self-gravity of a boosted polytrope with a median error below 1\% up to a Lorentz factor of 2, and keeps a binary of two polytropes stable over ten orbits. As an application, we study tidal disruption events (TDEs) of solar-type stars by supermassive BHs. The treatment of self-gravity shifts the disruption threshold and changes both the remnant core mass and the debris energy distribution. Newtonian self-gravity leaves substantial surviving cores in encounters that our prescription predicts to be nearly complete disruptions.
\end{abstract}


\section{Introduction}\label{sec:intro}
Smoothed particle hydrodynamics (SPH), introduced independently by \cite{Lucy1977} and \cite{GingoldMonaghan1977}, is a mesh-free Lagrangian method that represents a fluid by a set of particles and reconstructs its properties by kernel interpolation over neighbouring particles. Each particle carries a fixed mass, so the resolution adapts automatically to the density. This is suitable for systems that span a large range of densities, such as collapsing clouds or a star torn into debris streams. Because the method follows the fluid, it handles vacuum boundaries and arbitrary geometries without a mesh covering empty space. When the equations are derived from a Lagrangian, it also conserves mass, momentum, energy and angular momentum. Also, the particles are point masses, so the self-gravity of the fluid and the gravity of stars, dark matter or a central object can be computed together, with N-body methods. For these reasons SPH has become a central tool in computational astrophysics \citep{Monaghan1992,Monaghan1994,Monaghan2005}, from galaxy and cosmological structure formation \citep{HernquistKatz1989,Springel2005} and core-collapse supernovae \citep{FryerWarren2002} to star and planet formation, accretion discs, compact-object mergers and tidal disruption events (TDEs) \citep{Price2018}.

In almost all astrophysical applications, fluid self-gravity is as important as its hydrodynamics. In SPH, self gravity is usually computed in two parts. 
The force between nearby particles is softened with a gravitational kernel that is linked to the hydrodynamic smoothing kernel via the Poisson equation
\citep{PriceMonaghan2007}. The force from distant particles is computed by grouping them into clumps and using a multipole expansion of the resulting
gravitational potential, typically organised as a tree \citep{HernquistKatz1989}. Both are Newtonian and assume that the source moves slowly 
and that the gravitational field is weak. 

Many problems of current interest take place close to black holes (BHs) and neutron stars, where these assumptions may be challenged. Relativistic versions of SPH follow two approaches. In the first, the spacetime is evolved together with the matter, either by coupling SPH to a BSSN solver, as in SPHINCS\_BSSN \citep{RosswogDiener2021,Diener2022,Rosswog2023}, or by using the conformally flat approximation \citep{Oechslin2002,Bauswein2010}. Self-gravity is then part of the spacetime curvature. This is necessary when the fluid back-reacts on the spacetime, as in neutron star mergers, but is computationally expensive. The second approach applies when the fluid mass is much smaller than the mass of the central object, so that to a very good approximation, the fluid does not change the background spacetime. The metric is then held fixed and given analytically, for example by the Kerr or Schwarzschild solution, and the fluid is evolved on this background. General relativistic SPH (GRSPH) on such fixed backgrounds is now well established \citep{LiptaiPrice2019}. Because the background does not respond to the fluid, the fluid's own gravity has to be added separately, and in practice it is often included as a Newtonian source term in the momentum equation.

This Newtonian source term works well when the self-gravitating body moves slowly and stays far from the BH, beyond a few tens of gravitational radii. The gravity of the BH then changes over a scale much larger than the body, and self-gravity is a weak perturbation on the background. Strictly, however, the Newtonian potential is the weak-field metric perturbation in the rest frame of the source. Using it directly in the global frame, in which the source moves, is no longer correct when the source moves at a sizeable fraction of the speed of light or passes within a few gravitational radii of the BH. Both special and general relativistic corrections to self-gravity then become important. A consistent treatment computes the self-gravity in the local rest frame of the source and transforms it to the global frame in which the hydrodynamic equations are solved. Grid-based codes use this for a star by solving the Poisson equation in the Fermi normal frame of the star and adding the resulting potential to the metric \citep{ChengEvans2013,Ryu2020b}. A comparable treatment within SPH has received less attention.

Doing this in SPH raises problems that do not appear in a grid-based code. A Fermi normal frame is attached to one reference worldline. This is suitable for a single compact star, but not a fluid that spreads over a large region, such as tidal debris or several interacting bodies. In SPH every particle is a source, so a local rest frame has to be defined at the position of each particle. A second problem comes from the kernels. The hydrodynamic equations evolve the conserved density, momentum and energy (or entropy) in the global frame, and the conserved density is computed efficiently only with a kernel built from global coordinates. 
Because the gravitational softening is usually tied to this hydrodynamic kernel, it is also defined in global coordinates, whereas the self-gravity 
(that it softens) is a rest-frame quantity.
A moving body is Lorentz contracted along its motion, so a kernel that is isotropic in global coordinates is not isotropic in the rest frame of the source. Finally, distant particles enter the tree as nodes that are themselves moving, so the field of each node must also be corrected for its motion and for the background metric at its position.

In this paper we develop an improved prescription for the self-gravity of a fluid in GRSPH on a fixed background that systematically addresses these problems. We treat the self-gravity as a local, linear perturbation of the background metric and evaluate the field of each particle in its own rest frame. The resulting force takes the form of the standard Newtonian self-gravity with three changes: a) a prefactor that carries the specific enthalpy and the time dilation of the receiving particle, b) a stretch matrix that maps coordinate separations into the rest frame of the source, and c) a softening length defined in each particle's own rest frame. All three are local operations on a particle or a tree node, so the prescription fits into an existing tree-based GRSPH code. The derivation does not rely on any property of a particular astrophysical system. It requires only that the self-gravity be weak, that the fluid mass be small compared with the BH mass, and that the self-gravitating body be small in size compared with its distance from the BH. It therefore applies to any self-gravitating flow on a fixed BH spacetime that meets these conditions. We validate it with two tests : a boosted polytrope in flat spacetime, and a binary of two polytropes.

We then apply the prescription to TDEs, which are a natural first application of this improved version of self-gravity. Recall that when a star passes close to a massive BH, the tidal field of the BH may overcome the self-gravity of the star and disrupt it \citep{Hills1975,Rees1988}. The outcome is set by the impact parameter $\beta \equiv r_{\rm t}/r_{\rm p}$, where $r_{\rm p}$ is the pericentre distance and $r_{\rm t} \simeq R_\star (M_{\rm BH}/M_\star)^{1/3}$ is the tidal radius. Encounters with $\beta$ above a critical value $\beta_{\rm c}$, which depends on the stellar structure, fully disrupt the star. Those with $\beta \ll 1$ leave it essentially intact, and intermediate values strip the outer layers and leave a surviving core. For a parabolic orbit, roughly half of the stripped debris is bound to the BH and returns to pericentre, powering a luminous flare. In partial disruptions the surviving core encodes the of the encounter via mass and spin \citep{GuillochonRamirezRuiz2013,Mainetti2017,CoughlinNixon2019,Miles2020}, and on an eccentric orbit it can be disrupted again on later passages. For BH masses of $10^7\,M_\odot$ and above, the tidal radius of a solar-type star is about ten gravitational radii or less. Near pericentre the star then moves at a large fraction of the speed of light. Whether the star is fully or partially disrupted, and how much mass the surviving core keeps, depends directly on the balance between the tidal field and the self-gravity. These are therefore the quantities that are most sensitive to how the self-gravity is treated.

The rest of the paper is organised as follows. Section~\ref{sec:grsph} introduces the formalism and derives the self-gravity prescription. Section~\ref{sec:validation} presents the validation tests. Section~\ref{TDEResults} applies the prescription to TDEs of solar-mass stars by supermassive BHs and compares the results with those of the Newtonian treatment. Section~\ref{sec:conclusions} discusses the limitations of the present approximation and directions for future work.

\section{A self-gravitating fluid on a fixed background}\label{sec:grsph}

\subsection{Formulation}\label{sec:formulation}

Our hydrodynamic approach follows the GRSPH framework of \citet{LiptaiPrice2019}. We model the fluid using a finite set of SPH particles, with 
each carrying a conserved density $\rhostar=\sqrt{-g}\,\rho\,\frac{U^{0}}{c}$,
where $\rho $ denotes the rest-mass density as measured in the local comoving frame of the fluid element, $g$ represents the determinant of the underlying spacetime metric, and $U^{0}$ is the time component of the particle four-velocity $U^{\mu} = \frac{dx^\mu}{d\tau}$, where $\tau$ is the proper time and $x^\mu = (ct,x^i)$. The coordinate velocities are defined as $v^{\mu} = \frac{dx^\mu}{dt} = \frac{dx^\mu}{d\tau} \frac{d\tau}{dt} = \tfrac{U^{\mu} c}{U^0}$.

The canonical momentum per unit rest mass is $p_{i}=\tfrac{w\,U_{i}}{c^2}$ with $w=c^2+u+\frac{P}{\rho}$,
where $u$ is the specific internal energy and $w$ represents the specific enthalpy (incorporating the rest-mass contribution). 
In this subsection, for the rest of the discussion, we will set the speed of light $c=1$ in order to avoid cluttering the notation, and 
will restore it at the end. In this notation, the evolution equation that governs the fluid dynamics is given as
\begin{eqnarray}
  \frac{\dd p_{i}}{\dd t}
= -\frac{1}{\rhostar}\,\frac{\partial}{\partial x^{i}}\!\left(\sqrt{-g}\,P\right)
\;+\;\frac{w}{2} \frac{U^\mu U^\nu}{U^0} \frac{\partial g_{\mu\nu}}{\partial x^{i}} + \frac{P}{2\rho U^0} g^{\mu\nu} \frac{\partial g_{\mu\nu}}{\partial x^{i}} \label{eq1} 
    = -\frac{1}{\rho U^0}\frac{\partial P}{\partial x^{i}} + \frac{wU^0 }{2}v^\mu v^\nu \frac{\partial g_{\mu\nu}}{\partial x^{i}}~.
  \label{eq:mom}
\end{eqnarray} 
The first term in the last expression is the pressure gradient, and the second term is the gravitational acceleration. 

Now we add the contribution from the fluid's self-gravity $h_{\mu\nu}$ to the static background $g_{\mu\nu}$ as a dynamic perturbation and write the total metric as,
\begin{equation}
  \bar{g}_{\mu\nu}=g_{\mu\nu}+h_{\mu\nu},
  \label{eq:metricsplit}
\end{equation}
and we consider up to first order in the perturbation $h_{\mu\nu}$.
For the total metric we have $\bar{U}^0 = \left(-\bar{g}_{\mu\nu}v^{\mu}v^{\nu}\right)^{-\frac{1}{2}}$, which can be expanded as,
\begin{eqnarray}
\bar{U}^0  =  U^0\left[1 + \frac{1}{2}({ U^0})^2h_{\mu\nu}v^{\mu}v^{\nu}+ {\mathcal O}(h^2)\right],~~
\frac{1}{\bar U^0} = \frac{1}{{ U}^0}\left[1 - \frac{1}{2}({ U^0})^2h_{\mu\nu}v^{\mu}v^{\nu}+ {\mathcal O}(h^2)\right]~,
\label{Uzero}
\end{eqnarray}
where we define ${ U}^0= \left(-{ g}_{\mu\nu}v^{\mu}v^{\nu}\right)^{-\frac{1}{2}}$.

Now, we think of a `source' particle $b$ that generates the gravitational field and a `target' particle $a$ that experiences the gravity of the source. It is particle $a$ for which the equation of motion is being solved. The scalar quantity $(U_a^0)^2h_{\mu \nu}^{b}v^{\mu}_{a}v^{\nu}_{a}=h_{\mu \nu}^{b}U^{\mu}_{a}U^{\nu}_{a}  $ captures the interaction of particle $a$ with particle $b$'s gravitational field. 
Being a scalar, $h_{\mu \nu}^{b}U^{\mu}_{a}U^{\nu}_{a}$ can be evaluated in any frame, and in particular, we choose to evaluate this in the frame of the source $b$, i.e., $h_{\mu \nu}^{b}U^{\mu}_{a}U^{\nu}_{a} = h_{\mu \nu}^{'b}U^{'\mu}_{a}U^{'\nu}_{a}$, where we denote the 
rest frame quantities with a prime. In the rest frame of source $b$, we can write in standard notation, $h_{\mu\nu}^{'b}=-2\Phi_{b}\,\mathrm{diag}(1,1,1,1)$, where $\Phi_{b}$ is the Newtonian gravitational potential sourced by $b$. Then, using the normalisation of
the four velocity, we obtain the local quantity evaluated at the position of particle $a$ as
\begin{equation}
h_{\mu \nu}^{'b}U^{'\mu}_{a}U^{'\nu}_{a} = -2\Phi_{b}\left[2(U^{'0}_{a})^2 -1\right]~.
\label{sum1}
\end{equation}
Here, a prime denotes quantities measured in the frame of the source $b$, for example, the four velocity of $a$ is given by 
$U^{'\mu}_{a}=\left.\tfrac{dx^{'\mu}}{d\tau}\right |_{a}$. In this notation, $U^{'0}_{a}=\left.\tfrac{dx^{'0}}{d\tau}\right |_{a} = \gamma_{ab}$,
the Lorentz factor for $a$ as measured by $b$, and given by $\gamma_{ab} = - { g}_{\mu\nu}  U_{a}^\mu  U_{b}^\nu =  \Gamma_{a}  \Gamma_{b} \left(1 - \gamma_{ij}  V_{a}^i  V_{b}^j\right)$, with $ V_{a}^{i}$ the velocity component of $a$ as measured by a local Eulerian observer. Here, $\gamma_{ij}$ is the spatial metric and $\Gamma$ denotes the fluid Lorentz factor relative to the local Eulerian observer, both evaluated in the background spacetime. Now, for a given particle $a$, the net effect of all the source particles is given as a sum in our linear theory, and hence we obtain the effect of self-gravity as
\begin{equation}
h_{\mu\nu}U^{\mu}_{a}U^{\nu}_{a}=\sum_b h_{\mu\nu}^{'b}U^{'\mu}_{a}U^{'\nu}_{a}=-2\sum_b\Phi_{b}\left(2\gamma_{ab}^2 -1\right)\equiv -2\Psi_{a}~,
\end{equation}
where in the last step, we have introduced the generalised potential $\Psi \sim {\mathcal O}(h)$, 
which is a sum over the individual potentials of the same order appearing in Equation~(\ref{sum1}). It should be noted that each term in the summation is computed in different source frames. So, ideally, each term should be transformed into the rest frame of particle $a$ before the summation is taken. However, since $h_{\mu\nu}^{b}U^{\mu}_{a}U^{\nu}_{a}$ is an invariant quantity, it has the same value in any frame of reference, allowing us to add them directly. Now that we have a formula for the self-gravity acting on an individual particle, we can drop the particle label.

Next, substituting this result in Equation~(\ref{Uzero}), using the expression for $(U^0)^{-2}$ that also follows from this formula, and
consistently truncating the result beyond ${\mathcal O}(h)$, we finally obtain
\begin{equation}
\frac{1}{\bar U^0} = \frac{1}{{ U}^0}\left(1+\Psi\right)~,~~\bar U^0 = { U}^0\left(1-\Psi\right)~.
\end{equation}
Now we can finally substitute this result into Equation~(\ref{eq1}) and get the total momentum equation as

\begin{equation}
      \frac{\dd p_{i}}{\dd t}
= -\frac{1}{\rhostar}\,\frac{\partial}{\partial x^{i}}\!\left(\sqrt{-{g}}\,P\right)
\;+\;\frac{w}{2} \frac{{U}^\mu {U}^\nu}{{U}^0} \frac{\partial {g}_{\mu\nu}}{\partial x^{i}} + \frac{P}{2\rho {U}^0} {g}^{\mu\nu} \frac{\partial {g}_{\mu\nu}}{\partial x^{i}} + f^{\rm sg}_{i},
\label{eq:momentum_final}
\end{equation}
where $f^{\rm sg}_i$ is the acceleration due to self-gravity that we have obtained after subtracting the terms due to the background field from Equation~(\ref{eq1}) and can be written as
\begin{equation}
  f_{i}^{\rm sg}=\frac{w}{c^{2}}\left[-\frac{c}{{U}^{0}}\partial_{i}\Psi + \frac{{U}^{0}}{2c^3} \Psi\mathcal{D}_{i}\right] - \frac{\Psi\,\partial_{i}P}{\rho {U}^{0}c} + \frac{w}{c^3{U}^0}\Psi \partial_i \Psi,
  \label{eq:K}
\end{equation}
where $\mathcal{D}_{i}\equiv v^{\alpha}v^{\beta}\partial_{i}{g}_{\alpha\beta}$ is a purely background quantity representing the spatial gradient of the background metric contracted with the fluid velocity. Note that after restoring dimensions, ${U}^0 = c^2\left(-{g}_{\mu\nu} v^\mu v^\nu\right)^{-\frac{1}{2}}$. Also, $f_{i}^{\rm sg}$ is derived in the global frame in which the conservative momentum of the fluid is evolved. Similarly, $\partial_{i}\Psi$ is the derivative of $\Psi$ with respect to the global coordinates $x^i$. The method of calculating $\partial_{i}\Psi$ is discussed in the next subsection.

Now in Equation~(\ref{eq:K}), we can identify the first term as the leading order term. The third and fourth terms are suppressed by the stellar compactness, $\mathcal{C}_\star \equiv GM_\star/(R_\star c^2) \simeq 2.1\times10^{-6}$. Assuming $w\approx c^2$, the fourth term, $(w/c^3U^0)\,\Psi\,\partial_i\Psi$, differs from the leading term by a factor of $-\Psi/c^2$, giving a relative magnitude of $|\Psi|/c^2$. To estimate the third term, $-\Psi\,\partial_iP/(\rho U^0 c)$, we invoke hydrostatic equilibrium in the stellar rest frame, $\partial_iP/\rho=-\partial_i\Psi$. This relation remains valid in the coordinate frame because both gradients undergo the same spatial stretching. Consequently, the third term reduces to the fourth, and its relative magnitude is likewise $|\Psi|/c^2$. For a $1\,M_\odot$, $1\,R_\odot$ star, this ratio is approximately $2\times10^{-6}$ at the surface and remains below $10^{-5}$ at the centre, regardless of the orbit. Both terms can therefore be safely neglected. 

To assess the magnitude of the $\mathcal{D}_i$ term, we again consider a $1\,M_\odot$, $1\,R_\odot$ star orbiting a $10^6\,M_\odot$ Schwarzschild black hole at $r=6M$. Since $\mathcal{D}_\theta=\mathcal{D}_\phi=0$, this term contributes only in the radial direction. We therefore compare it with the radial component of the leading term, obtaining
\begin{equation}
  \frac{\left|\frac{U^0}{2c^3}\,\Psi\,\mathcal{D}_i\right|}
       {\left|\frac{c}{U^0}\,\partial_i\Psi\right|}
  = \frac{k}{2}\,\frac{(U^0/c)^2}{s_r}\,
  \frac{R_\star}{r}\,\frac{G M_{\rm BH}}{c^2 r},
  \label{eq:Dratio}
\end{equation}
where $s_r$ is the radial stretch between coordinate and rest-frame separations, and $k\equiv |\mathcal{D}_r|r^2/(GM_{\rm BH})$ is an orbit-dependent dimensionless quantity of order unity. The ratio is suppressed by the small stellar radius $R_\star$ relative to the black hole distance ($R_\star/r\approx 0.08$), and by $M/r=1/6$, while the remaining factors are of order unity. For radial infall, the ratio is $2.6\%$; at the pericentre of a parabolic orbit, it is approximately $1\%$. For a circular orbit, the ratio vanishes identically because $\mathcal{D}_r=0$ along any circular geodesic. 

Nevertheless, our TDE simulations show that including the $\mathcal{D}_i$ term makes a negligible difference to the dynamical outcome of a close encounter.
Because the pericenter crossing time $t_{\rm peri}\sim r_p/v_{\rm peri}\ll t_{\rm dyn}$, the short-duration radial impulse from the D-term integrates to a negligible change in the star's internal energy and bound mass fraction. All three sub-leading terms are therefore omitted. The self-gravity term finally is
\begin{equation}
  f_{i}^{\rm sg}\;\simeq\;-\frac{w}{{U}^{0}c}\,\partial_{i}\Psi\;=\;\frac{w}{c^{2}}\,\frac{{\alpha}}{{\Gamma}}\,a^{\rm sg}_i,
  \label{eq:fsg_impl}
\end{equation}
where ${\alpha}$ is the lapse function, and we define a new variable $a^{\rm sg}_i = -\partial_{i}\Psi$ as the bare-field acceleration.

\subsection{Implementation in smoothed particle hydrodynamics formalism}\label{sec:stretch}

While Equation~(\ref{eq:K}) derives the acceleration due to self-gravity, the prescription requires a functional form for $\Phi$. Since $\Phi$ is the gravitational potential defined in the rest frame where the Newtonian limit is valid, the separation between the source and target particles must be measured in the comoving frame of the source particle $b$. Traditionally, in the SPH formalism, distance is computed in global coordinates rather than proper distance. The conservative density and smoothing length associated with a particle are iteratively solved in the coordinate frame. This approach is convenient as it ensures that a uniform particle distribution yields a uniform conservative density. We follow the same convention from here on. 
Therefore, we construct the spatial vielbein $e^{\hat{i}}{}_{i}$ of the Eulerian observer using the spatial metric ${\gamma}_{ij}$:
\begin{equation}
  {\gamma}_{ij}=\delta_{\hat{i}\hat{j}}\,e^{\hat{i}}{}_{i}\,e^{\hat{j}}{}_{j}.
  \label{eq:triad}
\end{equation}
This vielbein is computed via Cholesky decomposition, which leaves an unimportant residual rotational freedom ($e\to Re$). Because any arbitrary rotation applied to the triad simultaneously rotates the boost vector $\hat{n}\to R\hat{n}$, the distance magnitude $|y|^{2}$ and the mapped gradient vector remain invariant under this gauge choice. A particle translating with a velocity ${V}^{i}$ relative to this Eulerian observer is reached via a Lorentz boost,
\begin{equation}
  {\Lambda}^{\hat{i}}{}_{\hat{j}}
  =\delta^{\hat{i}}_{\hat{j}}+\left(\Gamma-1\right)\hat{n}^{\hat{i}}\hat{n}_{\hat{j}},
  \qquad \hat{n}^i=\frac{{V^i}}{|{V}|}.
  \label{eq:lambda}
\end{equation}
Composing the vielbein and the boost generates a transformation that maps any global coordinate separation $\Delta x^{j}_{ab} = x^{j}_{a} - x^{j}_{b}$ into the orthogonal, comoving rest frame of the source:
\begin{equation}
  y^{\hat{i}}_{ab}={M_{b}}^{\hat{i}}{}_{j}\,\Delta x^{j}_{ab},
  \qquad {M_{b}}^{\hat{i}}{}_{j}={\Lambda_b}^{\hat{i}}{}_{\hat{j}}\,{e_b}^{\hat{j}}{}_{j}.
  \label{eq:M}
\end{equation}
We denote $M$ as the `stretch matrix'. 
Once mapped to the rest frame of $b$, the potential is evaluated directly at the distance $|y_{ab}|$. The corresponding gradient in the global coordinate system is extracted using the chain rule:
\begin{equation}
  \frac{\partial\Phi_b}{\partial (\Delta x^{i}_{ab})}
  =\frac{\partial y^{\hat{i}}_{ab}}{\partial (\Delta x^{i}_{ab})}\,\frac{\partial\Phi_{b}}{\partial y^{\hat{i}}_{ab}}
  = {M_{b}}^{\hat{i}}{}_{i}\,\partial_{\hat{i}}\Phi_{b}\!\left(|y_{ab}|\right),
  \label{eq:transpose}
\end{equation}
Now, discarding the subdominant corrections and expanding the spatial gradient, we arrive at the finalized analytic prescription:
\begin{equation}
  {f_{a}^{\rm sg}}_i
  = -K_{a}
    \sum_{b}(2\mathcal{\gamma}^2_{ab}-1)\,{M_{b}}^{\hat{i}}{}_{i}\,\partial_{\hat{i}}\Phi_{b}\!\left(|y_{ab}|\right) ,
  \label{eq:final}
\end{equation}
where the prefactor is defined as:
\begin{equation}
  K_{a} \equiv \frac{w_{a}{\alpha}_{a}}{c^{2}\Gamma_{a}}.
  \label{eq:Kdef}
\end{equation}
where $a$ and $b$ are indices for the receiver and source, respectively. Here, $\partial_{\hat{i}}\Phi_{b}$ is the negative of the Newtonian gravitational acceleration computed in the rest frame of the source particle $b$, transformed to the global frame using $M_{b}$, and the relative motion between the source and target particles contributes to the factor $(2\mathcal{\gamma}^2_{ab}-1)$. The prefactor $K_{a}$ enables the gravitational acceleration acting on the particle $a$ to be appropriately added as a source term in the conservative momentum evolution equation. 

Now, the final task is to compute the summation in Equation~(\ref{eq:final}) in the SPH formalism. Following \citep{Price2018}, the gravitational potential at particle $a$ is written as
\begin{equation}
\Phi_b(|y_{ab}|) = -\frac{G m_b}{2} \left[\phi(|y_{ab}|,\epsilon_{a}) + \phi(|y_{ab}|,\epsilon_{b}) \right].
\end{equation}
Here, $\epsilon_{a}$ is the softening length associated with the gravitational kernel $\phi$ in the rest frame of particle $a$. Note that $\Phi_{a}$ is a Newtonian quantity (not defined in the global frame). However, we need to transform the self-gravity acceleration acting on particle $a$ in the global frame to be used with Equation~(\ref{eq:momentum_final}).
The spatially adaptive smoothing and softening lengths, and energy conservation require that Equation~(\ref{eq:final}) be written as
\begin{equation}
\begin{aligned}
{f_{a}^{\rm sg}}_i
= -K_{a}\frac{G}{2}\sum_b m_b \left[\phi'(|y_{ab}|,\epsilon_{a}) + \phi'(|y_{ab}|, \epsilon_{b})\right] (2\mathcal{\gamma}^2_{ab}-1) \hat{y}_{\hat i} {M_{b}}^{\hat i}{}_{i} \\
- K_{a}\frac{G}{2}\sum_b m_b \left[ \frac{\zeta_a}{\Omega_a} \frac{\partial W(r_{ab}, h_a)}{\partial \boldsymbol{r_a}} + \frac{\zeta_b}{\Omega_b} \frac{\partial W(r_{ab},h_b)}{\partial \boldsymbol{r_a}} \right].
\label{eq:final_sph}
\end{aligned} 
\end{equation}
where $\phi' = -\tfrac{\partial\phi}{\partial |y_{ab}|}$
The unit vector $\hat y_{\hat i} = \delta_{\hat i \hat j} y^{\hat j}_{ab}/|y_{ab}|$. Also note that $\hat{y}_{\hat i} {M_{b}}^{\hat i}{}_{i}$ in matrix notation is $M^T \boldsymbol{\hat y}$. Also, $\boldsymbol{r_a}$ and $\boldsymbol{r_b}$ are global spatial coordinates of particle $a$ and $b$, respectively, and $r_{ab} = |\boldsymbol{r_a} - \boldsymbol{r_b}|$. The $\zeta$ and $\Omega$ are terms arising from variable smoothing length $h$ and variable softening length $\epsilon$, given as
\begin{equation}
\Omega_a = 1 - \frac{\partial h_{a}}{\partial \rho^*_{a}} \sum_b m_b \frac{\partial W(r_{ab},h_{a})} {\partial h_{a}},
\end{equation}
\begin{equation}
\zeta_a = \left(\det M_{a}\right)^{1/3}\frac{\partial h_{a}}{\partial \rho^*_{a}} \sum_b m_b (2\mathcal{\gamma}^2_{ab}-1) \frac{\partial \phi(y_{ab},\epsilon_{a})} {\partial \epsilon_{a}}.
\end{equation} 
Note that in SPH, it is customary to take the same value for the gravitational softening length and the hydrodynamic smoothing length since both are linked through the Poisson equation \citep{PriceMonaghan2007,SpringelHernquist2002}. Because $\phi$ is a rest-frame quantity evaluated at rest-frame distances, its softening length must also have a smoothing length counterpart defined in the rest-frame. Therefore, we define a rest-frame equivalent quantity of smoothing length $h$, namely the primitive smoothing length $h_{p}$. The physical volume element transforms via $\dd^{3}y=\det M\,\dd^{3}x$. Knowing that $\det\Lambda=\Gamma$ and $\det e=\sqrt{\gamma}$ with $\gamma = \det \gamma_{ij}$, we have
\begin{equation}
  \det M=\Gamma\sqrt{\gamma},
  \qquad
  h_{p}=h\left(\det M\right)^{1/3}.
  \label{eq:hp}
\end{equation}
The same scaling emerges from considering the physical densities as well. The proper rest-mass density is $\rho=\rho^*/\Gamma\sqrt{\gamma}$, so that in the rest frame, $h_{p}=\eta(m/\rho)^{1/3}={(\Gamma\sqrt{\gamma})}^{1/3}h$. Now, we can set the softening length $\epsilon = h_p$. With this choice, we ensure that energy conservation in the Newtonian limit is preserved. Note that Equation~(\ref{eq:final_sph}) is computed in the global frame. However, the self-gravity terms involving $\phi$ and $\zeta$ are rest-frame quantities. Therefore, after computing these terms in the rest frame, we transform them back to the global frame. The other terms involving $W$ and $\Omega$ are already defined in the global frame. 

When $|y_{(ab)}|$ is larger than the gravitational kernel size, $\phi$ and $\phi'$ take the usual Newtonian form, i.e., $\phi \propto 1/|y_{(ab)}|$ and $\phi' \propto 1/|y_{(ab)}|^2$. Equation~(\ref{eq:final_sph}) takes the form:
\begin{equation}
\begin{aligned}
\left. {f_{a}^{\rm sg}}_i \right|_{\rm far}
= -K_{a}G\sum_b \frac{m_b}{|y_{ab}|^2}(2\mathcal{\gamma}^2_{ab}-1) \hat{y}_{\hat i} {M_{b}}^{\hat i}{}_{i}
\label{eq:final_sph_far}
\end{aligned} 
\end{equation}
Now, in smoothed particle hydrodynamics, far-field gravity is not always calculated using every particle pair. Instead, a group of particles (a `node'), far from the receiver `leaf node' (the smallest possible node that includes the receiver particle $a$), is treated as a single entity with mass and quadrupole or higher-order moments,  \citep[see][]{Price2018}. The same prescription described above can be used to compute far-field gravity using Equation~(\ref{eq:final_sph_far}), for which we need to use the stretched distance between the centre of mass of the source node B and the receiver leaf node $A$, where the stretch matrix $M_B$ is computed in the rest frame of the source node $B$ using its centre of mass velocity. The relative velocity between the two nodes is used to find $\gamma_{AB}$. Once the acceleration at the centre of mass of leaf node $A$ is computed in the global frame, a Taylor expansion is used to find the acceleration at the position of particle $a$.     

It is important to discuss the validity of our approach here. The isotropic hydrodynamic kernel, defined in the global frame, does not remain isotropic in the rest frame due to relativistic length contraction (or rather, stretching). The spherical volume that contains a certain number of neighbours in the global frame takes an ellipsoidal shape in the rest frame to contain exactly those neighbours. Therefore, self-gravity due to those neighbours, being a rest-frame quantity, requires an ellipsoidal kernel. We bypass this issue by using a volume invariant spherical kernel with a primitive smoothing length $h_p$ that preserves the proper density of the particle. Our approach does not consider the exact same set of neighbour particles as in the global frame. This introduces an error when a strong density gradient is present. We argue that self-gravity, being a long-range force, accounts for all particles in the computational domain. As a result, the error in choosing a spherical kernel instead of its ellipsoidal shape can only cause a near-field gravity computation to be misclassified as a far-field gravity computation or vice versa. At higher density (higher resolution), the primitive smoothing length $h_p$ is small, thereby keeping the error negligible. On the other hand, at low densities (low resolution) where $h_p$ is large, the low total mass in that low-density region keeps the error low. 

A similar assumption is used in computing the gravitational kernel $\phi(|y_{ab}|,\epsilon_{a})$, where $|y_{ab}|$ is computed in the rest frame of particle $b$, whereas $\epsilon_{a}$ is computed in the rest frame of particle $a$. Relaxing this assumption requires one to compute the ellipsoidal shape of the gravitational kernel of particle $a$ in the frame of particle $b$. However, since strong near-field self-gravity appears in high-density regions, the relative velocity between two particles there is, in general, non-relativistic. In the case where two neighbouring particles are moving away from each other at relativistic velocities, this constitutes a diverging flow that automatically weakens self-gravity. It should be noted that this assumption is restricted to the near-field self-gravity only. For far-field self-gravity, the distance between a group of particles and the particle $a$ is computed in the frame of the source mass, correctly incorporating the relativistic length contraction. Therefore, our near-field self-gravity prescription is expected to improve the accuracy of remnant core mass calculation for partial tidal disruption events. Also, the far-field self-gravity prescription should accurately reproduce the orbital dynamics of a binary star system.

In the subsequent sections, whenever we refer to the Newtonian approach, we mean that the self-gravity is treated as purely Newtonian while retaining the GRSPH algorithm otherwise. Specifically, in the Newtonian approach, the self-gravitational potential is calculated in the unstretched global frame, and the prefactor $K_a=1$. Similarly, whenever we refer to the global frame transformed (GFT) self-gravity, we mean that
the self-gravity is evaluated using the full Lorentz-covariant formulation derived above, so that the
potential is evaluated in the stretched rest frame of the source via the matrix $M$
of Equation~(\ref{eq:M}), the softening length is set to the primitive value $h_{p}$ of
Equation~(\ref{eq:hp}), and the prefactor is $K_{a}=w_{a}\alpha_{a}/(c^{2}\Gamma_{a})$ of
Equation~(\ref{eq:Kdef}).

\section{Validation of the self-gravity treatment}\label{sec:validation}
We validate the prescription with two tests, each designed to isolate a different part of it. The first is a polytrope boosted to a uniform velocity in flat spacetime (Section~\ref{sec:criterion}). Its relativistic equilibrium is known exactly, so it tests the transformation of the self-gravity between the rest frame of the star and the global coordinate frame, through the prefactor $K_a$, the stretch matrix $M$ and the primitive softening length $h_p$, against an analytic solution. We perform this both for single force evaluation(static case) and over an extended evolution. The second is a binary of two polytropes in the Newtonian limit (Section~\ref{sec:binary}), which tests the far-field part of the gravity calculation and its conservation properties over several orbits. All simulations in the validation section and the TDE simulations in subsequent Section~\ref{TDEResults} were performed using our homegrown GRSPH code~\cite{Banerjee2026} based on the prescription of~\cite{LiptaiPrice2019}. The essential numerical tests establishing the robustness of our code are detailed in the Appendix of ~\cite{Banerjee2026}.
\subsection{Boosted polytrope test}\label{sec:criterion}
A star in hydrostatic equilibrium that moves uniformly through flat spacetime is an exact solution of the relativistic fluid equations, obtained from the static star by a Lorentz boost. It therefore provides a strict test of the transformation properties of the self-gravity, analogous to the boosted TOV-star test of \citet{Font2000}. We relax a polytropic star of mass $M_\star=1\,M_\odot$, radius $R_\star=1\,R_\odot$ and  adiabatic index $\gamma_{\rm ad}=5/3$ consisting of  $10^5$ SPH particles and boost it with Lorentz factor $\Gamma$. 

Under the Lorentz boost, the star is contracted by a factor $1/\Gamma$ along the direction of motion, while the conserved density measured in the global coordinate frame increases by a factor $\Gamma$. Here, proper density, pressure, and specific enthalpy, all of which are defined in the fluid rest frame, remain unchanged. The quantities evaluated before the application of the boost are denoted with a prime. The GFT prescription must therefore satisfy $\rhostar/\Gamma{\rhostar}'=h_{p}/h'=a_{\parallel}/\Gamma a_{\parallel}{}'=a_{\perp}/a_{\perp}{}'=1$
for all $\Gamma$, whereas $a_{\parallel}{}$ and $a_{\perp}$ are the components of bare-field acceleration ($\textbf{a}^{\rm sg}$) along the boost and perpendicular to the boost, respectively. Any departure from these relations, or from rigid translation, is an error of the self-gravity prescription or of the particle discretization. 
The Newtonian approach evaluates the potential on the contracted coordinates and, therefore, computes the field of an oblate spheroid instead of that of the spherical star in its rest frame. 

\begin{figure*}[t]
\includegraphics[width=\textwidth]{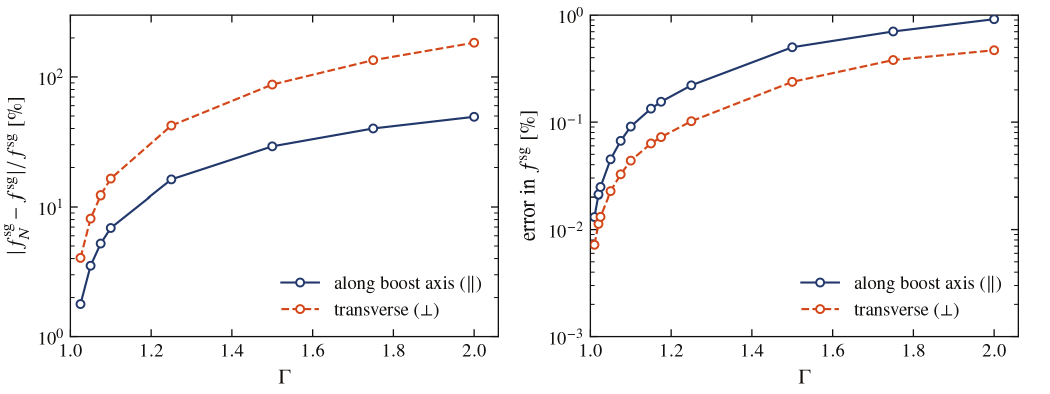}
\caption{Accuracy of the GFT self-gravity prescription for the boosted polytrope as a function of the Lorentz factor $\Gamma$. \textbf{Left Panel:} Median relative error in the assembled self-gravity force with respect to the analytical boosted target, shown along the boost direction ($\parallel$) and transverse to it ($\perp$). \textbf{Right Panel:} Median percentage difference between the self-gravity force obtained with the Newtonian and GFT self-gravity along the boost and transverse directions.}
\label{fig:accuracy}
\end{figure*}

Figure~\ref{fig:accuracy} shows the accuracy of the GFT self-gravity prescription for the boosted polytrope. The left panel shows the error in the assembled self-gravity force for the GFT treatment, where the error is calculated as the median over the particles of the relative difference between the simulated self-gravity force and the analytical prediction, following the procedure described in Appendix~\ref{app:grav_err}. The error remains under $1\%$ over the range of Lorentz factors considered. The right panel shows the percentage difference between the self-gravity force obtained with the GFT prescription and that obtained using the Newtonian treatment. The close agreement between the numerical results and these analytical predictions provides a useful check of the expected behaviour of the self-gravity calculation under a Lorentz boost.

The conventional Newtonian prescription fails to satisfy this condition, producing spurious internal accelerations in the boosted configuration. In contrast, our GFT approach preserves the expected force balance, demonstrating its consistency with the Lorentz transformation of the system.

\begin{figure*}[t]
\includegraphics[width=\textwidth]{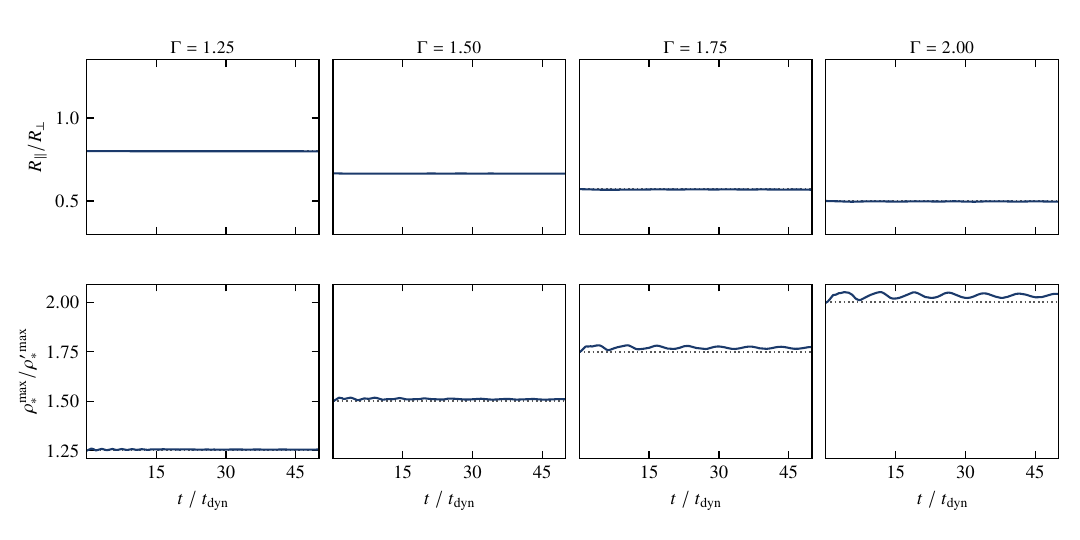}
\caption{Dynamical evolution of the boosted polytrope using the GFT self-gravity prescription for different Lorentz factors $\Gamma$. \textbf{Top Row:} Mass-weighted spatial axis ratio $R_{\parallel}/R_{\perp}$ as a function of time, with the analytical target $1/\Gamma$ shown by the dashed line. \textbf{Bottom Row:} Central conserved density $\rho^{*\max}$ normalized by its rest-frame value $\rho^{*\prime\max}$ as a function of time, with the analytical prediction $\rho^{*\max}/\rho^{*\prime\max}=\Gamma$ shown by the dotted line.}
\label{fig:ladder}
\end{figure*}

\begin{figure*}[t]
\includegraphics[width=\textwidth]{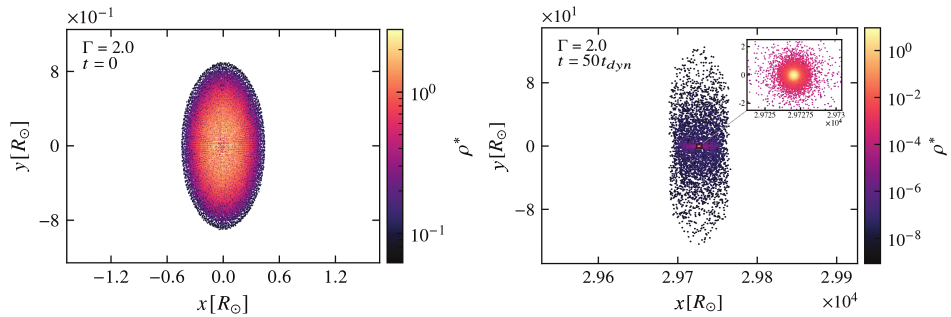}
\caption{Evolution of the boosted polytrope using the Newtonian approach for a boost factor $\Gamma=2$. \textbf{Left Panel:} Initial configuration of the boosted polytrope at the beginning of the simulation. \textbf{Right Panel:} Configuration after $50\,t_{\rm dyn}$. Self-gravity enhances due to wrong force calculation resulting in collapse of the star into a dense core. The inset shows the central dense core in greater detail.}
\label{fig:snap_newtonian}
\end{figure*}

The static test probes the force balance at a single instant. To test whether this accuracy is preserved during evolution, we evolve the boosted star for $(\Gamma\in\{1.25,1.50,1.75,2.00\})$ till $t=50\,t_{\rm dyn}$, with both treatments. The dynamical timescale $t_{\rm{dyn}}$ is defined as $\sqrt{\frac{R^3_\odot}{G M_\odot}}$ where $G=1, M_\odot=1, R_\odot=1$. The star should translate rigidly, with axis ratio $R_{\parallel}/R_{\perp}=1/\Gamma$ and a central conserved density $\Gamma$ times its rest-frame value (Figure~\ref{fig:ladder}). With the GFT prescription, Figure~\ref{fig:ladder} plots the axis ratios and ratios of central conserved density to the rest-frame density in the top row and bottom row, respectively. The dynamical evolution confirms that the static force accuracy aligns with the long term stability of the star.
The corresponding percentage errors from the analytical prediction are been shown in Figure~\ref{fig:ladder_err} in the Appendix \ref{appD2}.
Figures~\ref{fig:snap_fixed} and~\ref{fig:snap_newtonian} show the initial and final configurations for the highly boosted case, $\Gamma=2$, with the GFT approach and the Newtonian approach, respectively. In the boosted configuration, the GFT prescription preserves the internal state of the star within small numerical deviations.
The total energy is calculated as the sum of the kinetic energy in the centre-of-mass (CoM) frame, the internal energy, and the gravitational potential energy. The maximum relative energy change, measured with respect to the energy before the boost, remains in the range $10^{-3}$--$10^{-2}$. The final energy, evaluated at the last step of the evolution, differs from the initial energy by less than $0.2\%$. The internal velocity dispersion, defined as the root-mean-square (RMS) difference between particle velocities and the COM velocity, remains of the order $10^{-5}c$.
These results further support the stability of the GFT self-gravity prescription for a boosted equilibrium configuration.
With the GFT approach, the stretched polytrope retains its overall shape and density distribution after $50\,t_{\rm dyn}$, consistent with the expected rigidly translating equilibrium. These results provide a direct visual demonstration of the importance of the relativistic transformation of stellar self-gravity in the boosted configuration.

\begin{figure*}
\includegraphics[width=\textwidth]{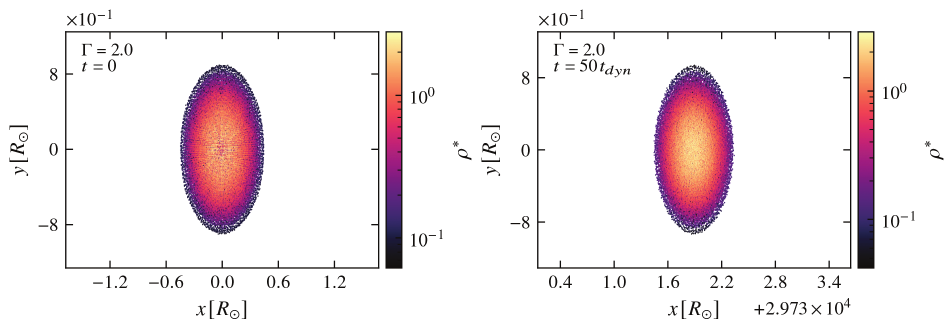}
\caption{Evolution of the boosted polytrope using the GFT self-gravity prescription for a boost factor $\Gamma=2$. \textbf{Left Panel:} Initial configuration of the boosted polytrope at the beginning of the simulation, showing the expected Lorentz-contracted shape and density distribution. \textbf{Right Panel:} Configuration after $50\,t_{\rm dyn}$, showing that the star retains its overall shape and density distribution during the evolution.}
\label{fig:snap_fixed}
\end{figure*}

\subsection{Binary polytrope}\label{sec:binary}
In our self-gravity treatment within GRSPH, the binary polytrope orbit problem provides a particularly significant test case.  When the separation between the two polytropes is much larger than their characteristic radii, the self-gravity calculation is dominated by the far-field gravity calculation. Additionally, the orbital dynamics provide a sensitive test of whether the perturbative metric generated by one star is correctly transmitted to the location of the other. 
\begin{figure*}[t]
\includegraphics[width=0.96\textwidth]{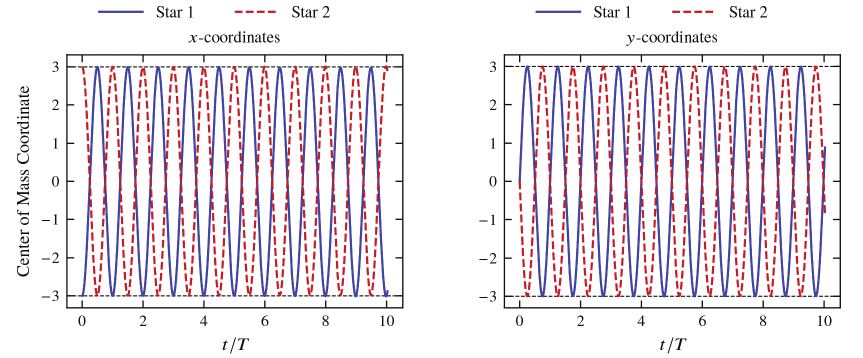}
\caption{Orbital evolution of the binary polytrope system over 10 orbital periods. \textbf{Left Panel:} $x$-coordinate of the center of mass of each star, measured with respect to the center of mass of the binary. \textbf{Right Panel:} $y$-coordinate of the center of mass of each star, measured in the same frame. The two stars remain on a nearly periodic orbit throughout the evolution, with an initial separation of $6\,R_\odot$.}
\label{fig:Binary}
\end{figure*}
In the weak-field limit, the resulting evolution should converge to the expected Newtonian binary dynamics, while maintaining the equilibrium structure of each star. More importantly, because the orbital motion integrates the gravitational acceleration over several dynamical times, small errors or inconsistencies in the perturbative self-gravity accumulate as secular errors in the orbital phase, separation, energy, momentum, and angular momentum. Thus, the method's ability to maintain a stable binary orbit over several orbital periods provides a sensitive test of whether the metric perturbation-induced self-gravity couples correctly to the GRSPH equations. We take two $\gamma_{\rm ad} = 5/3$ polytropic stars of $1\,M_\odot$ and $1\,R_\odot$  and initialize them on a circular orbit with $6\, R_\odot$ initial separation. Both stars have $20000$ particles each. The orbit is tracked for 10 orbital period(period denoted by $T$). The center-of-mass of both stars are plotted in Figure~\ref{fig:Binary}. The relative drifts (normalised) in conserved energy, momentum and angular momentum are of the order $10^{-6}$, $0.1$ and $0.01$, respectively. A relatively large momentum drift over 10 orbital periods is an expected consequence of the asymmetry in the far-field gravity.


\section{Comparative Study of Self-Gravity treatment in TDE simulations}\label{TDEResults}
We  compare the TDE outcomes of GFT prescription against the Newtonian approach, using identical initial conditions and numerical parameters, so that any difference arises from the treatment of self-gravity alone. We perform this set of simulations for $M_{\rm BH}=10^{6}$, $10^{7}$, and $2\times10^{7}\,M_\odot$ respectively for parabolic orbit at $15\, r_t$ on the $x$-axis (BH lies at the origin). We use a polytropic star of $\gamma_{\rm ad}=5/3$ and for each black hole mass we use the penetration factor $\beta=0.8$ and $1.0$.  
\begin{figure*}
    \centering
    \includegraphics[width=\textwidth]{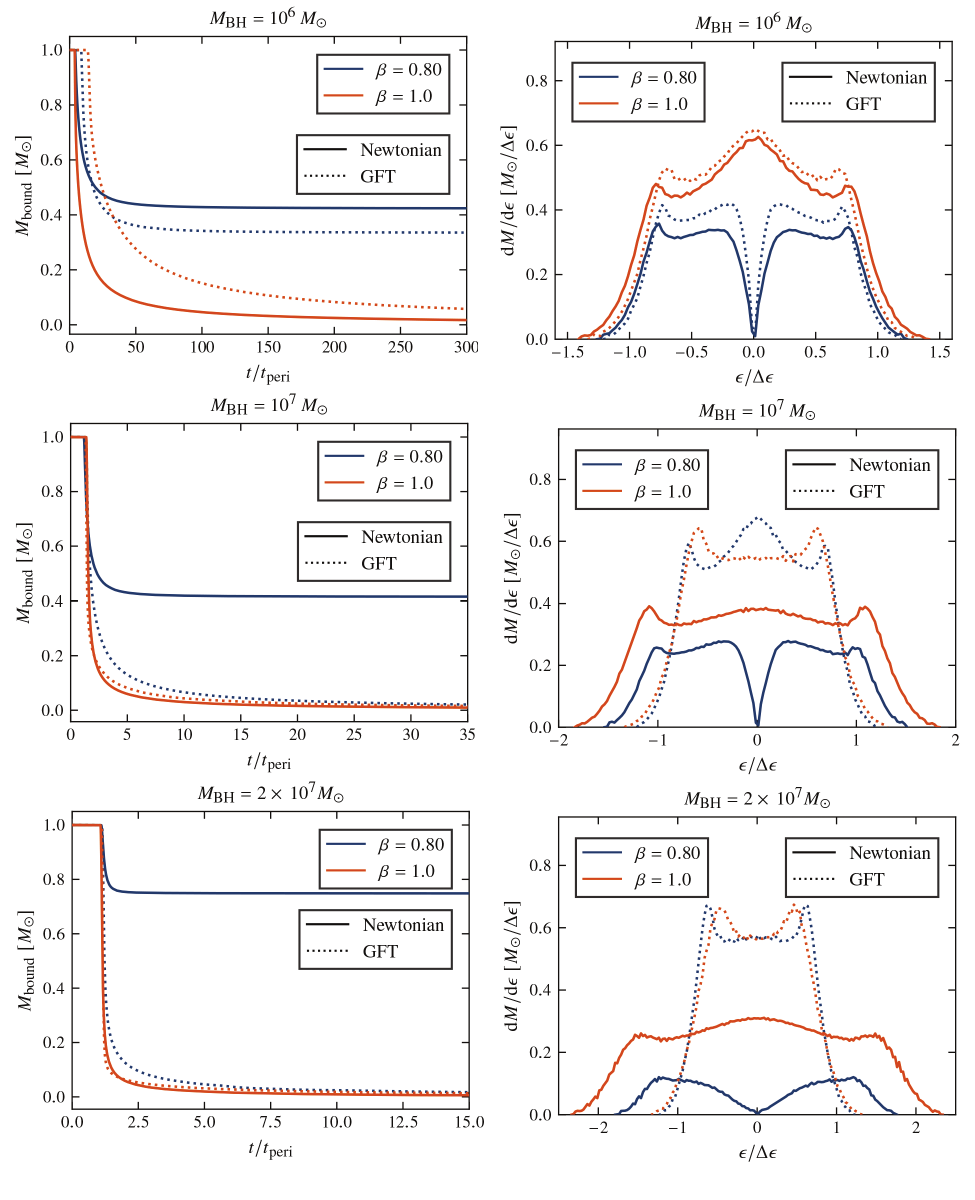}
    \caption{Evolution of the remnant core mass, $M_{\rm bound}$  with the pericentre normalised time, $t/t_{\rm peri}$ (\textbf{Left Panels}) and  distributions of debris energy, $dM/d{\epsilon}$ (\textbf{Right Panels}) from TDE simulations of  $M_{\rm BH}=10^{6}\,M_\odot$, $10^{7}\,M_\odot$, and $2\times10^{7}\,M_\odot$ . For $10^{6}\,M_\odot$, the remnant masses obtained with both the Newtonian and GFT approaches produce a similar surviving core mass whereas at higher black hole masses the Newtonian approach yields significantly larger $M_{\rm bound}$. The differences in energy spread increase with the blackhole mass. 
    }\label{fig:combined}
\end{figure*}
Figure~\ref{fig:combined} shows the $M_{\rm bound}$ and the debris mass distribution in specific orbital energy, $dM/d\epsilon$, for both approaches. $dM/d\epsilon$ in each case is plotted at a very late time post-pericenter passage when the specific orbital energies, $\Delta\epsilon$, exhibit no change with time evolution. Our simulations with the GFT approach qualitatively match with hydrodynamical studies \citep{Gafton2015,Ryu2020d}, the maximum gravity model introduced by \cite{CoughlinNixon2022} and the subsequent studies by \citep{Fancher2025, Bandopadhyay2024, Bandopadhyay2026}.
In encounters near the disruption threshold, the Newtonian prescription overestimates the self-gravity, creating a surviving core, whereas the GFT approach yields partial disruption with much lower remnant mass or complete disruption. In encounter scenarios with higher mass black holes, the Newtonian prescription may yield little more than a slight peeling off of the stellar surface or even a flyby with an almost intact star across all $r_p$ values. More importantly, the differences are not confined to encounters with a surviving core. Even in cases of complete disruption, the two prescriptions produce visibly different $dM/d\epsilon$ distributions. The difference in energy spreads grows with black hole mass, making the choice of self-gravity prescription more important for black holes with masses above $10^7\,M_\odot$.
Since this distribution determines the amount of debris returning to the black hole as a function of time, such differences can propagate into the predicted mass fallback rate, including the timing and magnitude of the peak fallback rate. Thus, the choice of self-gravity prescription can affect both the classification of the disruption outcome and the predicted debris return, with potential consequences for interpreting the observable emission from TDEs.

\section{Discussion and conclusions}\label{sec:conclusions}
We have developed a perturbative treatment of stellar self-gravity for fixed-background GRSPH simulations of tidal disruption events, as an alternative to the Newtonian source term that is commonly added to the momentum equation. The self-gravity of the star is treated as a first-order perturbation $h_{\mu\nu}$ of the background metric, and it enters the momentum equation through the scalar $h_{\mu\nu}U^{\mu}U^{\nu}$. Because this scalar is invariant, the contribution of each source can be evaluated in the rest frame of that source, where $h_{\mu\nu}$ can be written in terms of the Newtonian gravitational potential, and the contributions can be summed directly in the global frame in which the fluid equations are solved. In the SPH formalism, this leads to three modifications of the standard self-gravity calculation: a prefactor $K_{a}=w_{a}\alpha_{a}/(c^{2}\Gamma_{a})$, which is the same $w/(U^{0}c^2)$ factor that multiplies the pressure and background-gravity terms; a stretch matrix $M=\Lambda e$, which maps each coordinate separation into the rest frame of the source; and a primitive softening length $h_{p}=h\,(\Gamma\sqrt{\gamma})^{1/3}$, which matches the gravitational softening to the rest frame particle spacing. All three are local operations on a particle or a tree node, so our self-gravity prescription can be incorporated into an existing tree-based GRSPH code without altering its overall structure significantly.
The GFT prescription recovers the self-gravity of a boosted polytrope with a median error below $1\%$ up to $\Gamma=2$, and keeps the axis ratio and central density of the boosted star within $3\%$ of their equilibrium values over $50\,t_{\rm dyn}$. The Newtonian approach overestimates the same force by tens of percent already at $\Gamma=1.25$ and by up to $\approx185\%$ at $\Gamma=2$, and the star evolved with it collapses. The prescription also maintains a stable binary orbit over ten orbital periods.

The prefactor alone has a direct consequence for relativistic TDE simulations. For a parabolic orbit around a non-rotating black hole, $U^{0}/c=(1-2r_{g}/r_{p})^{-1}$ at pericenter. The Newtonian approach therefore overestimates the self-gravity there by $2r_{g}/(r_{p}-2r_{g})$, a fraction that grows as $\beta M_{\rm BH}^{2/3}$ for a given star, even before any error in the geometry of the field is included. Even when the self-gravity calculation is performed entirely in the global coordinate frame, with correct prefactor, omitting the transformation particle separations and source quantities into the local rest frame introduces an error in the calculated self-gravitational interaction. This can be further substantiated by boosted polytrope test incorporating the Lorentz-factor prefactor alone. It fails to recover the correct relativistic internal force balance and  resulting error increases with the boost velocity. In Section~\ref{TDEResults}, the Newtonian approach leaves remnant cores of $\approx0.4$ and $\approx0.75\,M_\odot$ at $\beta=0.8$ around the two heavier Schwarzschild black holes of $10^{7}\,M_\odot$ and $2\times 10^{7}\,M_\odot$ mass, whereas the GFT prescription gives near-complete disruption. TDE simulations that use Newtonian self-gravity around black holes of $\gtrsim10^{7}\,M_\odot$ for close encounter study, are therefore likely to be biased towards core survival. Locating the boundary between partial and full disruption correctly also matters for interpreting observations because models of partial disruptions can reproduce the optical and UV properties of detected TDEs, and some events classified as full disruptions may in fact be partial \citep{Sharma2026a}.

The present approach has several limitations that define the scope of its applicability. It assumes a quasi-static, weak stellar self-gravity field and a stellar mass negligible compared with that of the black hole. Nor does it capture the dynamical back-reaction of the star on the background geometry as well, which may become important in encounters involving more compact objects.  
With these tests in place, our approach can be applied to surveys of partial TDEs across deep encounters, black hole masses, and spins, enabling the use of observable signatures, such as remnant masses, kick velocities, and debris fallback rates to constrain black hole properties and probe strong-gravity effects in relativistic encounters.

\section*{Acknowledgements}
We acknowledge the support and resources provided by PARAM Sanganak under the National Supercomputing Mission, Government of
India, at the Indian Institute of Technology Kanpur.

\section*{Data Availability}
The data underlying this article will be shared upon reasonable request of the corresponding author.

\bibliographystyle{mnras}

\appendix
\section{Explicit form of the spatial vielbein}\label{appA}

The spatial three-metric $\gamma_{ij}$ can be decomposed into an upper-triangular vielbein matrix $e^{\hat{k}}{}_{j}$ via the Cholesky decomposition, satisfying $\gamma_{ij}=\delta_{\hat{k}\hat{l}}e^{\hat{k}}{}_{i}e^{\hat{l}}{}_{j}$. This is mathematically equivalent to completing the square on the line element $\dd s^{2} = \gamma_{ij}\dd x^{i}\dd x^{j}$ sequentially for $\dd x^{1}$, $\dd x^{2}$, and $\dd x^{3}$. Writing this out in full for a general metric with six independent components yields exact, closed-form algebraic expressions for the vielbein components:
\begin{align}
  e^{\hat{1}}{}_{1} &= \sqrt{{\gamma}_{11}}, &
  e^{\hat{1}}{}_{2} &= \frac{{\gamma}_{12}}{e^{\hat{1}}{}_{1}}, &
  e^{\hat{1}}{}_{3} &= \frac{{\gamma}_{13}}{e^{\hat{1}}{}_{1}}, \nonumber \\
  e^{\hat{2}}{}_{1} &= 0, &
  e^{\hat{2}}{}_{2} &= \sqrt{{\gamma}_{22} - \left(e^{\hat{1}}{}_{2}\right)^{2}}, &
  e^{\hat{2}}{}_{3} &= \frac{{\gamma}_{23} - e^{\hat{1}}{}_{3}e^{\hat{1}}{}_{2}}{e^{\hat{2}}{}_{2}}, \label{eq:vielbein} \\
  e^{\hat{3}}{}_{1} &= 0, &
  e^{\hat{3}}{}_{2} &= 0, &
  e^{\hat{3}}{}_{3} &= \sqrt{{\gamma}_{33} - \left(e^{\hat{1}}{}_{3}\right)^{2} - \left(e^{\hat{2}}{}_{3}\right)^{2}}. \nonumber
\end{align}
Because $\gamma_{ij}$ is a proper spatial metric, it is strictly positive-definite. Consequently, the arguments inside every square root above are mathematically guaranteed to be positive for any valid spacetime geometry. This closed-form representation maps directly into the numerical implementation, ensuring an exact local frame extraction free from iterative root-finding.

\section{Error calculations}
\subsection{Error in assembled self-gravity}\label{app:grav_err}
For each particle, we quantify the relative deviation of the post-boost value from its target value, where the target is defined by that particle's corresponding pre-boost value rather than by a fitted analytical relation. Specifically, we calculate the absolute fractional deviation, $|\mathcal{q}_{\rm post}/\mathcal{q}_{\rm target}-1|$, for each particle. The analysis is restricted to the inner $90\%$ of the stellar mass, excluding surface particles, whose density estimates are inherently more susceptible to one-sided kernel sampling and associated noise, even in the correct treatment. We then take the median of the fractional deviations over the selected core particles and multiply it by $100$ to obtain the percentage error. By construction, a value of zero corresponds to perfect agreement with the target values.
\begin{equation}
  \varepsilon = 100 \times \operatorname{median}_{a\,\in\,\rm core}
  \left|\frac{q_{a}}{q_{a}^{\rm target}}-1\right|
  \label{eq:errstat}
\end{equation}
We report the median of the particle-wise absolute fractional deviations, rather than the absolute value of the median signed deviation. This measure quantifies the typical magnitude of the deviation without allowing positive and negative deviations to cancel. For measurement of the parallel component of assembled gravity, the target here is the pre-boost value, unchanged.

\begin{align}
  \varepsilon_{\parallel} = 100 \times \operatorname{median}_{a\,\in\,\rm core}
  \left|\frac{f^{\rm sg}_{\parallel}({\rm post})}
             {{f^{\rm sg}_{\parallel}}({\rm pre})}-1\right|,
\varepsilon_{\perp} &= 100 \times \operatorname{median}_{a\,\in\,\rm core}
  \left|\Gamma\,\frac{f^{\rm sg}_{\perp}({\rm post})}
                     {f^{\rm sg}_{\perp}({\rm pre})}-1\right|
\end{align}

The arguments, "post" and "pre" refer to the state after and before the applied boost in the first step of the simulation. For the perpendicular component of the assembled gravitational field, the target value is the pre-boost value divided by $\Gamma$. We therefore rescale the perpendicular component by $\Gamma$, so that the target value becomes unity, consistent with the parallel component.
\subsection{Dynamical errors of the boosted configuration}\label{appD2}
For the dynamical boosted-polytrope test, we also quantify the deviation of the evolved configuration from the expected Lorentz-contracted equilibrium. We consider the mass-weighted spatial axis ratio and the central conserved density, whose expected values are $1/\Gamma$ and $\Gamma\rho^{*\prime\max}$, respectively. The corresponding percentage errors are defined as
\begin{align}
\epsilon_R =
\left|
\frac{R_{\parallel}/R_{\perp}}{1/\Gamma}-1
\right|\times 100,
\epsilon_\rho &=
\left|
\frac{\rho_*^{\max}}{\Gamma\rho^{*\prime\max}}-1
\right|\times 100.
\end{align}
\begin{figure*}[t]
\includegraphics[width=\textwidth]{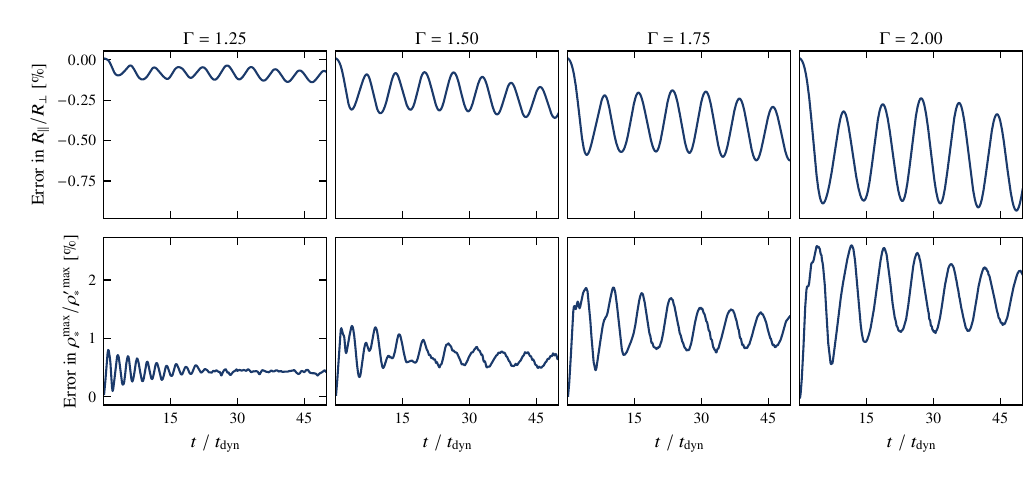}
\caption{Percentage error of the boosted-polytrope diagnostic quantities as a function of time $t / t_{\rm dyn}$ across different Lorentz factors $\Gamma$ for the GFT formulation. \textbf{Top row:} Percentage error in the mass-weighted axis ratio $R_\parallel / R_\perp$ relative to the theoretical prediction $1/\Gamma$. \textbf{Bottom row:} Percentage error in the central conserved density ratio $\rho^{*\max} / \rho^{*\prime\max}$ relative to the theoretical prediction $\Gamma$.}
\label{fig:ladder_err}
\end{figure*}
Figure~\ref{fig:ladder_err} shows these errors as a function of time for the different Lorentz factors considered. The axis-ratio and central-density errors remain small throughout the evolution, indicating that the GFT prescription maintains the expected Lorentz-contracted structure of the boosted configuration over dynamical timescales.
\label{lastpage}

\end{document}